\documentclass[10pt,aps,prb,twocolumn,showpacs,amsmath,amssymb,superscriptaddress]{revtex4-1}
\usepackage{graphicx}
\usepackage{natbib}
\usepackage{amsmath}
\usepackage{xcolor}
\usepackage{comment}
\usepackage{soul}
\usepackage{xfrac}
\usepackage{mathtools}
\usepackage[colorlinks=true, urlcolor  = blue, citecolor = blue, linkcolor = blue]{hyperref}

\begin{document}

\title{Switching between Magnetic Bloch and N\'eel Domain Walls with Anisotropy Modulations}

\author{K\'{e}vin J. A. Franke}
\affiliation{School of Physics and Astronomy, University of Leeds, Leeds LS2 9JT, United Kingdom}

\author{Colin Ophus}
\affiliation{National Center for Electron Microscopy, Molecular Foundry, Lawrence Berkeley National Laboratory, Berkeley, California 94720, USA}

\author{Andreas K. Schmid}
\affiliation{National Center for Electron Microscopy, Molecular Foundry, Lawrence Berkeley National Laboratory, Berkeley, California 94720, USA}

\author{Christopher H. Marrows}
\affiliation{School of Physics and Astronomy, University of Leeds, Leeds LS2 9JT, United Kingdom}

\date{\today}

\begin{abstract}
It has been shown previously that the presence of a Dzyaloshinskii-Moriya interaction in perpendicularly magnetized thin films stabilizes N\'eel type domain walls. We demonstrate, using micromagnetic simulations and analytical modeling, that the presence of a uniaxial in-plane magnetic anisotropy can also lead to the formation of N\'eel walls in the absence of a Dzyaloshinskii-Moriya interaction. It is possible to abruptly switch between Bloch and N\'eel walls via a small modulation of both the in-plane, but also the perpendicular magnetic anisotropy. This opens up a route towards electric field control of the domain wall type with small applied voltages through electric field controlled anisotropies. 
\end{abstract}
\maketitle

\begin{figure}
\centering
\includegraphics{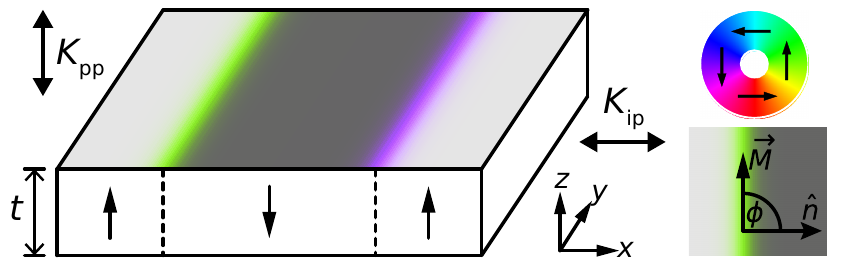}
\caption{\label{geometry} (left) Sketch of the simulation geometry with definition of directions. (right) Definition of in-plane color wheel and domain wall magnetization angle $\phi$ between the magnetization $\protect\overrightarrow{M}$  at the centre of the domain wall and the normal $\hat{n}$ to the domain wall.}
\end{figure}

\begin{figure}
\centering
\includegraphics{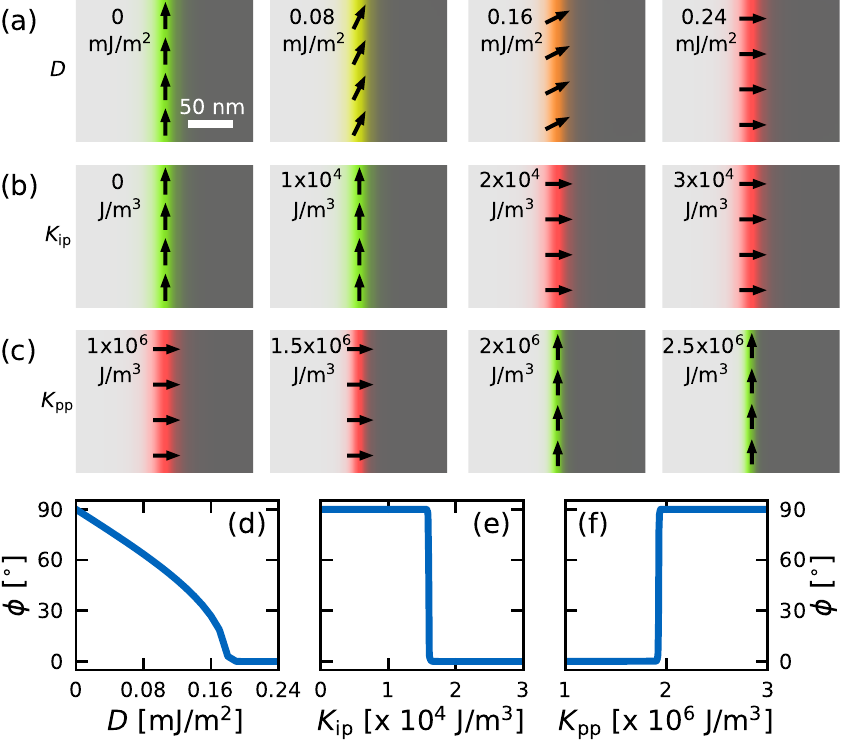}
\caption{\label{DMI-and-Ki} Domain wall images as a function of (a) DMI constant $D$ and (b) in-plane anisotropy constant $K_{\mathrm{ip}}$ for a perpendicular anisotropy constant $K_{\mathrm{pp}}=1\!\times\! 10^{6}$ J/m$^{3}$. (c) Images as a function of $K_{\mathrm{pp}}$ for $K_\mathrm{ip}=3\!\times\! 10^{4}$ J/m$^{3}$. Corresponding domain wall magnetization angles $\phi$ as a function of (d) $D$, (e) $K_{\mathrm{ip}}$, and (f) $K_{\mathrm{pp}}$.}
\end{figure}

\begin{figure}
\centering
\includegraphics{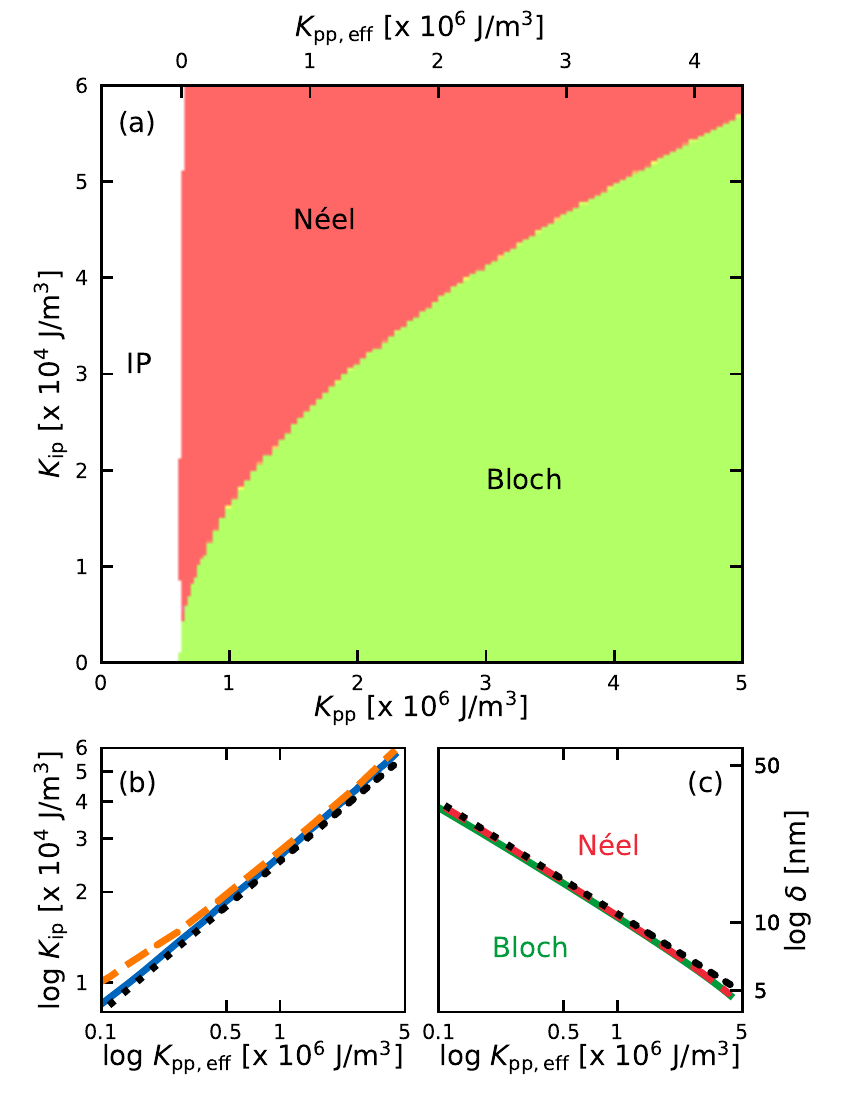}
\caption{\label{Ko-vs-Ki}(a) Phase diagram of the domain wall magnetization angle $\phi$ as a function of perpendicular ($K_{\mathrm{pp}}$) and in-plane ($K_{\mathrm{ip}}$) anisotropy constants. (b) Location of the transition between Bloch and N\'eel walls for a thin film (blue line), for a nanowire (orange dashed line), and according to the analytical model (black dotted line). (c) Widths $\delta$ of Bloch (green line) and N\'eel (red dashed line) domain walls just below and above the transition. The analytical expression $\delta=2\,\sqrt{\sfrac{A}{K_{\mathrm{pp,eff}}}}$ is shown as a black dotted line.}
\end{figure}

\begin{figure}
\centering
\includegraphics{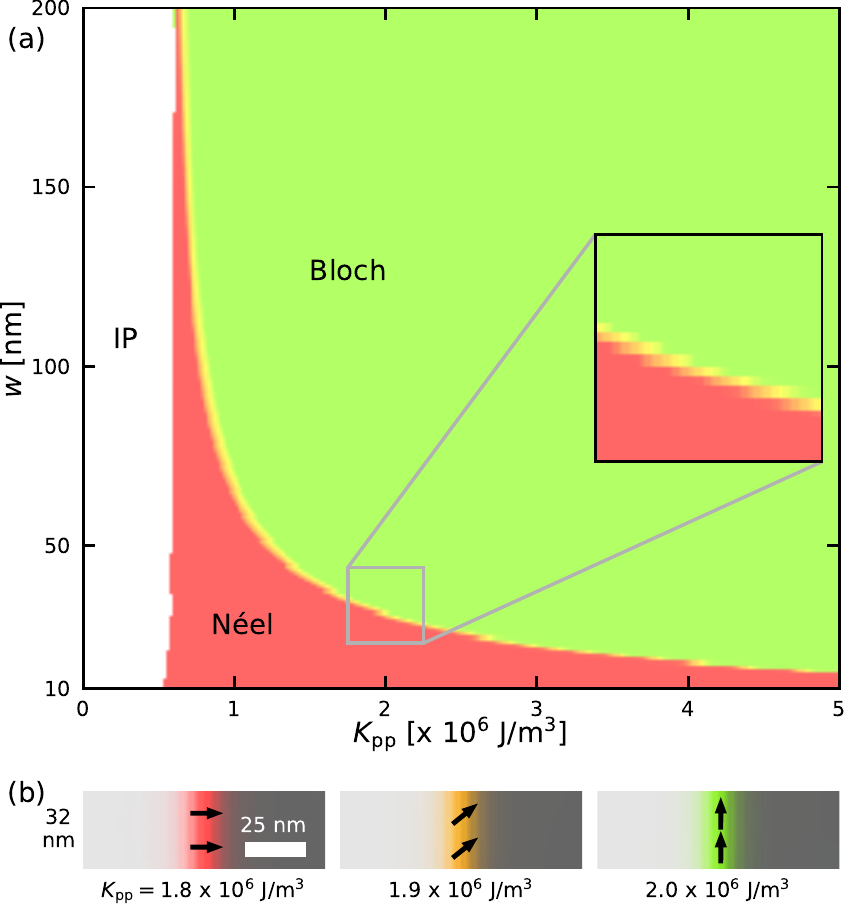}
\caption{\label{Ko-vs-w}(a) Phase diagram of the domain wall magnetization angle $\phi$ as a function of perpendicular magnetic anisotropy constant $K_{\mathrm{pp}}$ and nanowire width $w$. (b) Domain wall images as a function of $K_{\mathrm{pp}}$ for $w=32$ nm.}
\end{figure}

\begin{figure}
\centering
\includegraphics[width=1\linewidth]{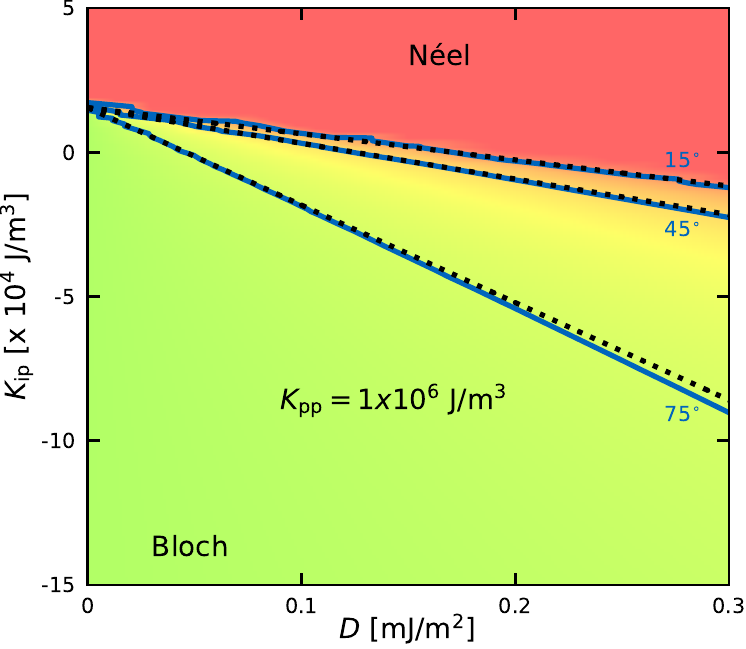}
\caption{\label{Ki-vs-DMI} Phase diagram of the domain wall magnetization angle $\phi$ as a function of DMI constant $D$ and in-plane anisotropy constant $K_{\mathrm{ip}}$ for a fixed magnitude of the perpendicular anisotropy constant $K_{\mathrm{pp}}=1\!\times\! 10^{6}$  J/m$^{3}$. Solid blue lines are contour lines of the simulation data for given domain wall magnetization angles of $15^{\circ}$, $45^{\circ}$, and $75^{\circ}$. Dotted black lines are the contour lines expected from the analytical model.}
\end{figure}

The presence of an interfacial Dzyaloshinskii-Moriya interaction (DMI) in perpendicular (denoted ``PP'') magnetized thin films stabilizes N\'eel type domain walls (DWs) of fixed chirality\cite{Thiaville_2012,chen_tailoring_2013} as opposed to the Bloch DWs favored by magnetostatics that are formed in the absence of a DMI. \cite{hubert_schafer_BOOK} In nanowires, Né\'eel DWs of fixed chirality have been shown to be driven efficiently in the same direction as the conventional electric current by interfacial spin-orbit torques,\cite{Ryu2013,emori_currentdriven_2013} making them appealing for potential DW devices. \cite{parkin_magnetic_2008} \\
At the same time, electric field control of magnetism holds the promise of low-power spintronic devices. Particularly the modulation of both in-plane (IP) and PP magnetic anisotropies is well-established. Control is achieved either via interfacial strain transfer from a ferroelectric or piezoelectric substrate and inverse magnetostriction, \cite{lahtinen_pattern_2011,streubel_strainmediated_2013,yu_straininduced_2015,shepley_modification_2015,finizio_magnetic_2014} or via direct charge modulation at the interface with an insulator. \cite{niranjan_electric_2010,maruyama_large_2009,shiota_quantitative_2011,Wang2012,bauer_magnetoionic_2015} The latter modulates the interface anisotropy, which arises from the broken translational symmetry at the interface and spin-orbit coupling (SOC), and can give rise to PP magnetic anisotropy (PMA). \cite{johnson_1996,aharoni_BOOK} \\
Broken spatial inversion symmetry and SOC are also the ingredients that give rise to the DMI. It emerges at the interface of a ferromagnet with a heavy metal, \cite{FertLevy1980} or more generally at the interface with a different material due to Rashba SOC, as a result of the electrostatic potential difference between the materials.\cite{Kundu2015} The latter induces a DMI at the interface between a ferromagnetic film and an insulator, and can thus be sensitive to a gate voltage. \cite{Nawaoka_2015,Srivastava2018,Zhang2018,Suwardy_2019,Yang2020,Schott_2021} This has been used for electric field control of magnetic DW motion via the modulation of the DMI.\cite{Koyama2018,Koyama_2020} The DMI has also been shown to be sensitive to the application of strain,\cite{Gusev_2020,Yang2020} which opens up the route towards electric field control of DMI via coupling to a piezoelectric or ferroelectric substrate. Still, both mechanisms for tuning the DMI will also affect the magnetic anisotropy, making it difficult to disentangle their effect on magnetic DWs. Furthermore, a switch between DW types (Bloch and N\'eel), or a reversal of chirality with voltage remains elusive. Similarly, the voltage control of skyrmions is currently being investigated, and the electric field induced creation, annihilation and even motion have been demonstrated. \cite{Hsu2017,Schott2017,Srivastava2018,Ma2019} As for the case of DWs, electric fields generally affect several material parameters, making it difficult to determine the mechanism that allows for this voltage control. \\
Recently, \citet{chen_unlocking_2015} reported that in a magnetic multilayer exhibiting PMA and DMI, the type of DW  depends on the relative angle between the DW and a uniaxial IP magnetic anisotropy (IMA) of constant magnitude. Given the strong dependence of spin-orbit torques on DW type and the fact that magnetic anisotropies can be induced and modulated in various ways, \cite{hubert_schafer_BOOK} this observation raises the question about control -- and possibly switching --  of DW type with anisotropy modulations. \\
In this letter, we therefore demonstrate an alternative mechanism for the control of DW type: using micromagnetic simulations and analytical modeling, we show that the presence of a uniaxial IMA of fixed orientation can also lead to the formation of N\'eel DWs in the absence of a DMI. It is possible to abruptly switch between Bloch and N\'eel DWs via a small modulation of the strength of both the IMA and PMA. This opens up a route towards efficient electric field control of the DW type with small applied voltages, as the magnetic anisotropy strength can be modulated via the direct voltage controlled magnetic anisotropy mechanism,\cite{maruyama_large_2009,shiota_quantitative_2011} or via magnetoelastic anisotropy induced through coupling to a piezoelectric element.\cite{LiLarge2015,yu_straininduced_2015}  \\
We investigated this control of DW type through micromagnetic simulations using the OOMMF software package.\cite{donahue_oommf_1999} The simulated geometry is sketched in Fig.~\ref{geometry}: it consists of a thin film of thickness $t=1$ nm. 
The in-plane dimensions are $400 \times 200$ nm$^2$, and two-dimensional periodic boundary conditions \cite{wang_twodimensional_2010} are used to simulate an infinite film. Simulations are initialized such that two DWs are stabilized. We choose reasonable values for the saturation magnetization $M_{\mathrm{s}}=1\times 10^{6}$ A/m and exchange stiffness $A=3\times 10^{-11}$ J/m.\cite{stohr_Book,Eyrich2012,Devolder2016} We consider the effects of a PMA with anisotropy constant $K_{\mathrm{pp}}$, a uniaxial IMA along the $x$-direction (perpendicular to the DWs) with anisotropy constant $K_{\mathrm{ip}}$, and an interfacial DMI with constant $D$. \\
To simulate a nanowire geometry, the two-dimensional periodic boundary conditions are omitted and the width of the simulations altered in the $y$-direction. The extent of simulations in the $x$-direction is chosen such that DWs are not affected by finite size effects along this dimension. 
The DW magnetization angle $\phi$ is defined relative to the DW normal $\hat{n}$ (Fig.~\ref{geometry}). For Bloch DWs $\phi= \pm 90^{\circ}$, while for N\'eel DWs $\phi = 0^{\circ}$ or $180^{\circ}$. 
The DW width $\delta =\int_{-\infty}^{+\infty} \cos^{2}(\theta) dx$ is defined as an integral over the magnetization profile of the DW, where $\theta = \sin^{-1}\left(\sfrac{M_{\mathrm{z}}}{M_{\mathrm{}s}}\right)$ is the polar angle between the magnetization direction and the film plane. \cite{jakubovics_comments_1978} For an ideal Bloch DW this definition yields $\delta = 2\sqrt{\sfrac{A}{K}}$, where $K$ is the effective anisotropy. \cite{hubert_schafer_BOOK}\\ 
We start by reproducing the well-known effect the DMI has on the chirality of magnetic DWs in PP magnetized thin films. Images of a DW as a function of increasing DMI constant $D$ for $K_\mathrm{pp}=1\!\times\! 10^{6}$ J/m$^{3}$ are shown in Fig.~\ref{DMI-and-Ki}(a). The corresponding $\phi$ is plotted in Fig.~\ref{DMI-and-Ki}(d). As reported previously,\cite{Thiaville_2012} the DW magnetization angle rotates continuously from a Bloch towards a N\'eel configuration as soon as a DMI is present. Above a certain value of $D$, $\phi$ saturates at $0$, i.e.\ a N\'eel DW. \\
In the absence of a DMI, an IMA with easy axis perpendicular to the DW also allows for a tuning between Bloch and N\'eel DWs. The effect of an increasing IMA is shown in Fig.~\ref{DMI-and-Ki}(b). Unlike the DMI, the anisotropy does not immediately affect the DW magnetization angle. As a function of increasing $K_{\mathrm{ip}}$, the DW first remains of Bloch type until it switches abruptly to a N\'eel DW. This behaviour is highlighted in Fig.~\ref{DMI-and-Ki}(e), where $\phi$ is shown as a function of $K_{\mathrm{ip}}$. Note, that the magnitude of the IMA required to switch between DW types is about two orders of magnitude smaller than the PMA strength and does thus not significantly affect the magnetization in the domains. \\
For $K_\mathrm{pp}=1\!\times\! 10^{6}$ J/m$^{3}$ and $K_\mathrm{ip}=3\!\times\! 10^{4}$ J/m$^{3}$, a N\'eel DW is stabilized. As shown in the images of Fig.~\ref{DMI-and-Ki}(c), and the graph in panel (f), an increase in the PMA strength eventually leads to an abrupt switch to a Bloch DW. It is thus possible to switch between DW types by either tuning the IMA or PMA strength. 
We further investigate this in a phase diagram (Fig.~\ref{Ko-vs-Ki}(a)), establishing regions where N\'eel or Bloch DWs are stabilized as a function of $K_{\mathrm{pp}}$ and $K_{\mathrm{ip}}$. We find that for higher values of $K_{\mathrm{ip}}$ and lower values of $K_{\mathrm{pp}}$, N\'eel DWs form. Conversely, for smaller values of $K_{\mathrm{ip}}$ and larger values of $K_{\mathrm{pp}}$, Bloch DWs are observed. The transition between Bloch and N\'eel DWs appears sharp, which is in stark contrast to the continuous transition observed when the DMI constant is changed. A $K_{\mathrm{pp}}$ -- vs -- $D$ phase diagram in the Supplemental Information\cite{SupMat} (SI) furthermore reveals that in the presence of a DMI, but absence of IMA, a tuning of the magnitude of $K_{\mathrm{pp}}$ has no effect on the DW magnetization angle.
It is thus only this new mechanism, involving an IMA, that allows for switching between DW types via a modulation of the PMA strength, at least for the experimentaly achievable parameters considered here. \\
We investigate this surprising result further by plotting the IMA strength at which the transition occurs as a function of the effective PMA (blue line in Fig.~\ref{Ko-vs-Ki}(b). We find a linear dependence of $\log \left(K_{\mathrm{ip}}\right)$ on $\log \left(K_{\mathrm{pp,eff}}\right)$ with slope $s\!=\!\sfrac{1}{2}$. The IMA strength at which the switch between DW types occurs thus shows a square root dependence on the effective PMA strength. \\
To understand this dependence, we construct a simple analytical model. The full derivation can be found in the SI.\cite{SupMat} The model compares the total energies of Bloch and N\'eel DWs for a given magnitude of $K_{\mathrm{pp}}$ and $K_{\mathrm{ip}}$. The widths of both N\'eel and Bloch DWs are shown as a function of $K_{\mathrm{pp,eff}}$ in Fig.~\ref{Ko-vs-Ki}(c), along with the theoretical value $\delta=2\,\sqrt{\sfrac{A}{K_{\mathrm{pp,eff}}}}$. We find excellent agreement between them, and therefore make the simplification that both types of DW types exhibit the same width. \\
We find that in first approximation the difference in DW surface energy $\sigma$ between N\'eel ($\sigma_{\mathrm{N}}$) and Bloch ($\sigma_{\mathrm{B}}$) DWs is given by:
\begin{equation}
\Delta \sigma =\sigma_{\mathrm{N}} -\sigma_{\mathrm{B}} = K_{\mathrm{ip}}\, \delta - \frac{\mathrm{ln}2}{\pi}\, \mu_{0}\, M_{\mathrm{s}}^{2}\, t.
\label{DeltaEdens}
\end{equation}
The first term results from the IMA, while the second term is a consequence of magnetostatics. The magnetostatic contribution arises from magnetic volume charges only.\cite{SupMat,Skaugen_2019,garcia-cervera2004,Kohn2005}
For low values of $K_{\mathrm{ip}}$, the magnetostatic energy favoring Bloch DWs dominates. At large values of $K_{\mathrm{ip}}$, the anisotropy energy favoring N\'eel DWs overcomes the magetostatic energy. The transition between DW types is expected to occur when the difference in energy is zero. As a result, the DW is expected to switch between Bloch and N\'eel type when:
\begin{equation}
K_{\mathrm{ip}}= \frac{\mathrm{ln}2\, \mu_{0}\, }{2\, \pi} \frac{M_{\mathrm{s}}^{2}\, t}{\sqrt{A}} \sqrt{K_{\mathrm{pp,eff}}}.
\label{Kipswitch}
\end{equation}
This dependence is plotted in Fig.~\ref{Ko-vs-Ki}(b), where excellent agreement between results from micromagnetic simulations (blue line) and the analytical model (black dotted line) is observed. The analytical model thus explains the square root dependence of the IMA strength at which the switch between DW types occurs on the effective PMA strength. The fact that the type of DW that is stabilized also depends on $K_{\mathrm{pp}}$ is due to the fact that the IMA contribution in $\Delta \sigma$ (Eq.~\ref{DeltaEdens}) depends on the DW width, which in turn depends on the PMA. \\
Our simulations correspond to an experimental system where IP and PP magnetic anisotropies can be tuned independently. Electric field control of DW type could be achieved by tuning the strength of one of these anisotropies with a voltage. One way would be to deposit a magnetic multilayer exhibiting PMA onto a piezoelectric substrate to induce a voltage tuneable uniaxial IMA via interfacial strain transfer and inverse magnetostriction. \cite{finizio_magnetic_2014,LiLarge2015}
Another approach, that would also allow for local control, would be to tune the PMA strength via charge modulation at an interface. \cite{niranjan_electric_2010, maruyama_large_2009, shiota_quantitative_2011, Wang2012, bauer_magnetoionic_2015}
This would of course require the presence of an uniaxial IMA, which could be induced in various ways. 
One way to achieve this would be to simply utilize the shape anisotropy in a magnetic nanowire to induce a uniaxial IMA. DWs tend to form perpendicular to the nanowire length, while the shape anisotropy induces a uniaxial anisotropy along it, which corresponds to the geometry investigated here. Nanowires are used in most DW applications, and this approach would eliminate the need for a separate mechanism to induce the uniaxial IMA. \\
Fig.~\ref{Ko-vs-w}(a) displays the phase diagram for the DW type as a function of $K_{\mathrm{pp}}$ and nanowire width $w$. It confirms previous observations of a transition from a Bloch to a N\'eel DW when the nanowire width is reduced. \cite{Martinez2011,DeJong2015} 
It does also show that this transition depends on the strength of the PMA. Therefore, it is possible to switch between Bloch and N\'eel DWs for a given nanowire width when $K_{\mathrm{pp}}$ is modulated. We extract the location of the transition and express it in terms of effective anisotropies.\cite{SupMat} The resulting curve is plotted as an orange dashed line in Fig.~\ref{Ko-vs-Ki}(b). It matches the results for thin films and the analytical model well, except for low values of the effective IMA and PMA strengths. We ascribe this to the fact that for wide nanowires (corresponding to a low effective $K_{\mathrm{ip}}$), expressing magnetostatic effects as a simple uniaxial anisotropy is too crude an approximation. \\
Unlike the case of thin films, the transition between DW types in nanowires does not result from a competition between a magnetic anisotropy and magnetostatics. it is purely the results of magnetostatics: magnetic volume charges favor Bloch DWs, while magnetic surface charges on the edges of the nanowire are minimized for N\'eel DWs. For a given width of the nanowire, increasing $K_{\mathrm{pp}}$ decreases the width of the DW which leads to a reduction of magnetic surface charges. As a result, Bloch DWs become energetically favorable. Conversely, decreasing $K_{\mathrm{pp}}$ increases the build up of magnetic surface charges, thus favoring N\'eel DWs. \\
The nanowire geometry also allows for the stabilization of intermediate DW magnetization angles $\phi$. As highlighted by the inset in Fig.~\ref{Ko-vs-w}(a), the transition between Bloch and N\'eel DWs is not as sharp as in the thin film case. This has already been observed as a function of $w$.\cite{Boehm2017} As shown in Fig.~\ref{Ko-vs-w}(b), tuning between N\'eel and Bloch DWs with a PMA in nanowires also involves DWs with intermediate $\phi$. \\
Stabilizing N\'eel DWs with an IMA does not favor one chirality, unlike the DMI. Left- and right-handed DWs are energetically degenerate. For applications, it might be necessary to obtain N\'eel DWs with a fixed chirality. We now show that it is still possible to tune DW type with an anisotropy in the presence of a small DMI that yields DWs of fixed chirality. Phase diagrams as a function of $K_{\mathrm{pp}}$ and $K_{\mathrm{ip}}$ for fixed values of $D$ are shown in the SI.\cite{SupMat} Here, we focus on the phase diagram as a function of $D$ and $K_{\mathrm{ip}}$ for a fixed value of $K_{\mathrm{pp}}=1\times10^6$ J/m$^3$ shown in Fig.~\ref{Ki-vs-DMI}. 
We observe that while a positive value of $K_{\mathrm{ip}}$ can be used to switch from a Bloch to a N\'eel DW when $D=0$, a negative value of $K_{\mathrm{ip}}$ tunes the N\'eel DW obtained for large values of $D$ towards a Bloch DW. For values of $D$, where an intermediate $\phi$ is obtained, negative and positive $K_{\mathrm{ip}}$ values tune the DW towards the Bloch and N\'eel type, respectively. A negative $K_{\mathrm{ip}}$ corresponds to an easy axis along the DW. We observe furthermore that while the transition between Bloch and N\'eel DWs is abrupt for $D=0$, it becomes increasingly wider as $D$ increases. This is highlighted by the contour lines (blue) for DW magnetization angles of $15^{\circ}$, $45^{\circ}$, and $75^{\circ}$. \\
The contour lines can be obtained from our analytical model by including the DMI energy in the DW surface energy. It yields the black dotted lines in Fig.~\ref{Ki-vs-DMI}, showing excellent agreement between micromagnetic simulations and the model, and demonstrating that the contour lines are linear in $-D$.\cite{SupMat} \\
We have therefore shown, using micromagnetic simulations, that the presence of a uniaxial in-plane magnetic anisotropy can lead to the formation of N\'eel domain walls in the absence of a DMI. It is possible to abruptly switch between Bloch and N\'eel walls via a small modulation of not only the in-plane, but also the perpendicular magnetic anisotropy. In nanowires, the shape anisotropy can be used to induce the in-plane anisotropy. In this case, tuning between domain wall types with a perpendicular magnetic anisotropy proceeds via intermediate domain wall magnetization angles. The presence of a DMI widens the transition between domain wall types. A simple analytical model accounts for the dependence of domain wall type on both the in-plane and perpendicular magnetic anisotropies, and the DMI.  Our results open up the route towards voltage control of domain wall type with small applied voltages through electric field controlled anisotropies. As only N\'eel domain walls are driven by interfacial spin orbit torques in nanowires, while Bloch domain walls are not, this could allow for efficient control of doman wall motion with electric fields. We expect that our results obtained for DWs can be extended to other chiral spin textures such as skyrmions.

\begin{acknowledgments}
This project has received funding from the European Union’s Horizon 2020 research and innovation programme under the Marie Sklodowska-Curie grant agreement No 750147. K.J.A.F. acknowledges support from the Jane and Aatos Erkko Foundation. Work at the Molecular Foundry was supported by the Office of Science, Office of Basic Energy Sciences, of the U.S. Department of Energy under Contract No. DE-AC02-05CH11231. This research used the Lawrencium computational cluster resource provided by the IT Division at the Lawrence Berkeley National Laboratory (Supported by the Director, Office of Science, Office of Basic Energy Sciences, of the U.S. Department of Energy under Contract No. DE-AC02-05CH11231.

\end{acknowledgments}


%

\end{document}